\documentclass[12pt, onecolumn, draftcls]{IEEEtran}
\usepackage{epsfig, psfrag, amsmath, amssymb, amsfonts, latexsym, eucal, balance, hyperref, indentfirst, graphicx, bookmark}
\usepackage[tight,footnotesize]{subfigure}
\newtheorem{thm}{Theorem}

\newtheorem{lem}{Lemma}

\newtheorem{defn}{Definition}

\begin{document}

\title{Non-cooperative Game For Capacity Offload}
\author{Feng Zhang, Wenyi Zhang, \textit{Senior Member, IEEE}, and Qiang Ling, \textit{Senior Member, IEEE}
\thanks{FZ and WZ are with Department of Electronic Engineering and Information Science, University of Science and Technology of China, Hefei 230027, China (email: zhangfg@mail.ustc.edu.cn, wenyizha@ustc.edu.cn); QL is with Department of Automation, University of Science and Technology of China, Hefei 230027, China (email: qling@ustc.edu.cn). The research of FZ and WZ has been supported by MIIT of China through grants 2010ZX03003-002 and 2011ZX03001-006-01.
}
}

\maketitle
\thispagestyle{empty}

\begin{abstract}
With the blasting increase of wireless data traffic, incumbent wireless service providers (WSPs) face critical challenges in provisioning spectrum resource. Given the permission of unlicensed access to TV white spaces, WSPs can alleviate their burden by exploiting the concept of ``capacity offload'' to transfer part of their traffic load to unlicensed spectrum. For such use cases, a central problem is for WSPs to coexist with others, since all of them may access the unlicensed spectrum without coordination thus interfering each other. Game theory provides tools for predicting the behavior of WSPs, and we formulate the coexistence problem under the framework of non-cooperative games as a capacity offload game (COG). We show that a COG always possesses at least one pure-strategy Nash equilibrium (NE), and does not have any mixed-strategy NE. The analysis provides a full characterization of the structure of the NEs in two-player COGs. When the game is played repeatedly and each WSP individually updates its strategy based on its best-response function, the resulting process forms a best-response dynamic. We establish that, for two-player COGs, alternating-move best-response dynamics always converge to an NE, while simultaneous-move best-response dynamics does not always converge to an NE when multiple NEs exist. When there are more than two players in a COG, if the network configuration satisfies certain conditions so that the resulting best-response dynamics become linear, both simultaneous-move and alternating-move best-response dynamics are guaranteed to converge to the unique NE.
\end{abstract}

\newpage
\setcounter{page}{1}

\section{\textsc{Introduction}}
\label{sect:introduction}

With the blasting increase of wireless data traffic from new applications such as smartphones, a solution for wireless service providers (WSPs) is to make use of the concept of ``capacity offload'' to transfer part of their traffic load elsewhere off the main system, to alleviate the load thus improving the overall system capacity. In industry, femtocell and WiFi networks are the primary candidates for capacity offload \cite{meshkati2009mobility} \cite{roh2011femtocell}. As the Federal Communications Commission (FCC) and other regulatory bodies have recently permitted unlicensed access to TV white space spectrum \cite{fcc2010rules}, it becomes possible for WSPs to use dynamic spectrum access techniques to offload their traffic from their own licensed bands (i.e., private bands) to the publicly available unlicensed bands (i.e., shared bands). A natural question thus comes to us: what is the impact of the additional unlicensed spectrum on coexistence? A general drawback associated with unlicensed spectrum is the so-called tragedy of the commons \cite{nguyen2011impact}, i.e., the spectrum may be overused by WSPs without admission fee and the communication may encounter excessive interference \cite{peha2009sharing}.

\begin{figure}[!ht]
\centering
\subfigure[System model]{
\label{fig:model}
\includegraphics[width=4in]{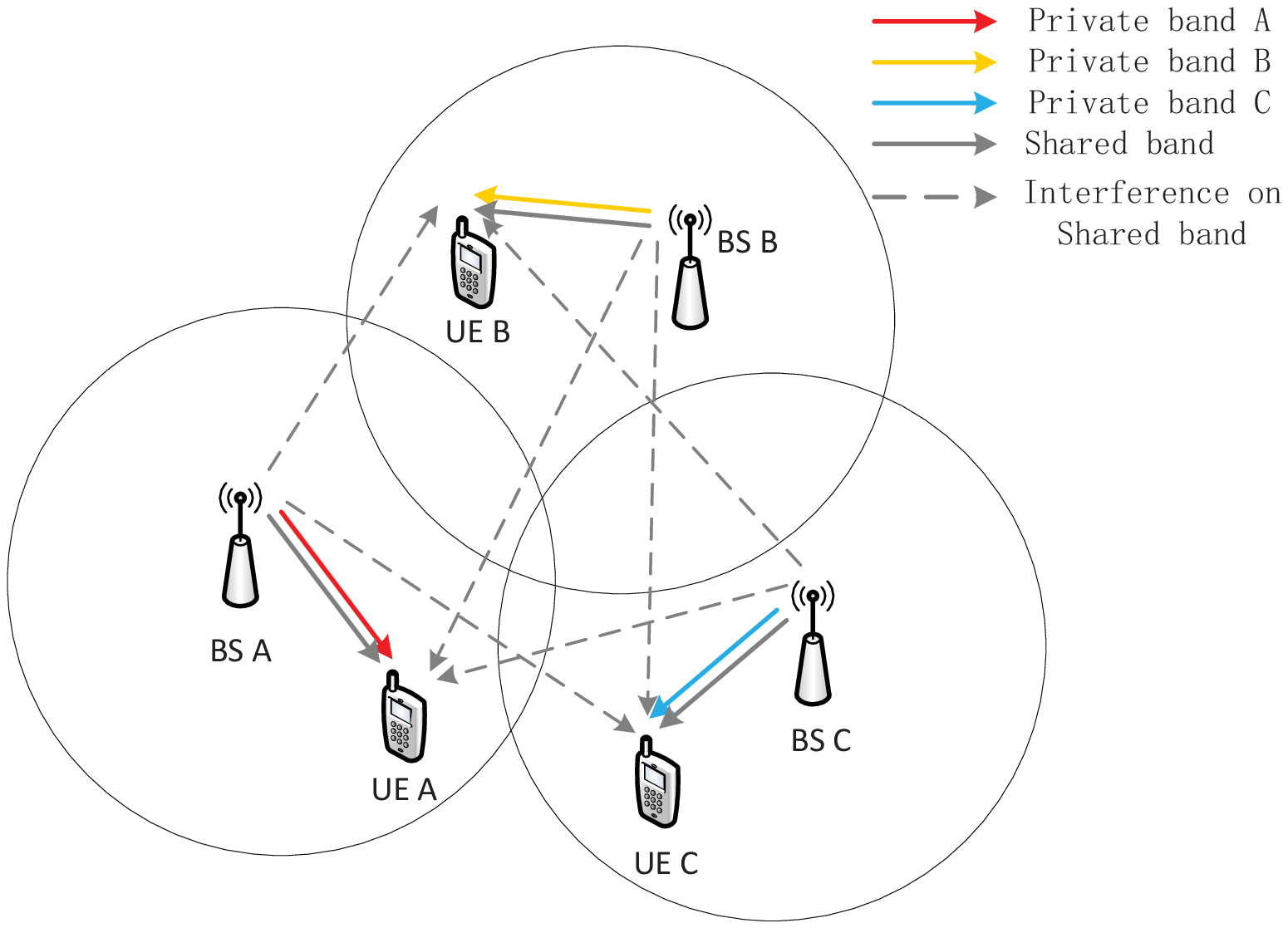}}
\subfigure[Spectrum allocation scheme]{
\label{fig:spectral}
\includegraphics[width=2in]{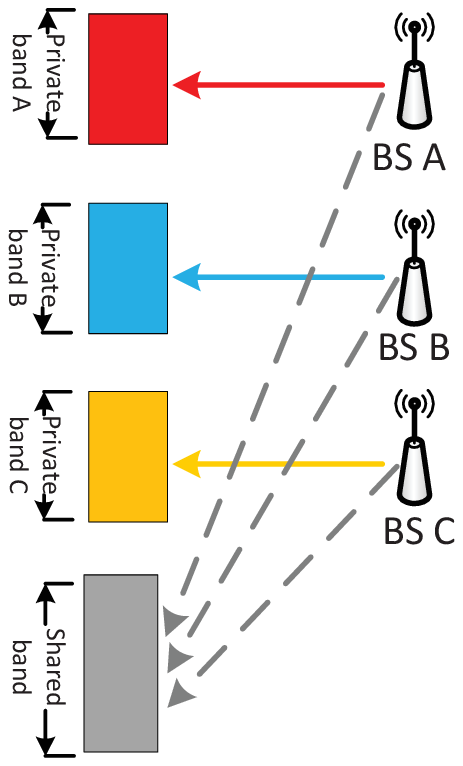}
}
\caption{An example of typical capacity offload scenarios}
\label{fig:models}
\end{figure}

We consider a simplified network model as depicted in Figure \ref{fig:models}. Suppose a time-division channel access scheme, so that in each time slot only one user equipment is active in each WSP's network. The WSPs utilize the additional unlicensed spectrum simultaneously, due to absence of coordination among different WSPs on spectrum allocation. A user equipment can communicate with its serving WSP in the corresponding private band and the shared band at the same time. Communication in the private band is free from interference of other WSPs, while in the shared band, a receiver will experience interferences from other WSPs' transmissions. A transmitter has an average power constraint and thus needs to allocate its power budget between its private band and the shared band. The power allocation strategies of WSPs thus lead to an inherent interaction among the WSPs, thus determining their behaviors.

Game theory is a powerful tool in analyzing and predicting outcomes of interactive decision making processes. In this work, we use game theory to analyze the interaction among WSPs, and formulate such interaction into a capacity offload game (COG): WSPs are players of the game, power allocation schemes (i.e., how to split each WSP's power budget between private band and shared band) constitute the strategy space of each WSP, and achievable rates are the utility functions of WSPs. Due to the lack of coordination, the relationship among the WSPs is competitive; that is, a COG is non-cooperative and each WSP would choose a power allocation strategy to attempt to maximize its achievable rate regardless of other WSPs.

There has recently been a heightened interest in the applications of game theory in wireless networks; see, e.g., \cite{han11:book} and references therein. Works whose models bear similarities to ours, however, are relatively few, summarized as follows. In \cite{belmega2009resource}, the authors considered a network consisting of two interfering links for which both sources have access to a common relay which has access to bands orthogonal to that used by the sources. In \cite{mochaourab2009resource}, the authors considered a network consisting of two source-destination links and three bands, assuming that one of the bands is shared by the transmitters while the other two bands are private. The analysis therein is based on results of supermodular games, with strategy spaces defined in a way such that the game has strategic complementarities. Unfortunately, that approach does not apply for games with more than two players, as that we consider in this paper.

Through the analysis of the COG, we arrive at a number of interesting conclusions. The COG always has at least one pure-strategy NE when all WSPs adopt deterministic strategies (i.e., pure strategies). When WSPs reach an NE, none of them would have incentive to unilaterally deviate from the NE since otherwise the deviation would decrease its utility \cite{nash1950equilibrium}. Even if we permit mixed strategies, i.e., an WSP choosing its strategy randomly according to a probability distribution over its strategy space, the COG does not possess any mixed-strategy NE. The NE of a COG is not necessarily unique, and the number of NEs depends upon the network parameters. As an illustration, we fully characterize the NEs for two-player COGs, for all possible network parameters.

We then examine the behavior of best-response learning algorithms \cite{fudenberg1998theory} which requires only local information for each WSP, to study how to reach an NE in a distributed way. In best-response learning algorithms, a COG is played repeatedly and in each time slot, an WSP updates its strategy based on its best-response function with respect to other WSPs' strategies in the previous time slot. WSPs can update their strategies either simultaneously or alternatingly. Our findings are as follows. For a two-player COG, alternating-move best-response dynamics always converge to an NE, regardless of the number of NEs in the game; whereas the simultaneous-move best-response dynamic does not always converge to an NE when multiple NEs exist, depending upon the initial strategies adopted. For COGs with more than two WSPs, the convergence property is generally difficult to analyze due to the nonlinearity in the best-response functions. But we find that when the network parameters are configured in a way such that each WSP's best-response function is linear with respect to other WSPs' strategies, both simultaneous-move and alternating-move best-response dynamics are guaranteed to converge to the unique NE of the COG.

The remaining part of the paper is organized as follows. Section \ref{sect:models} formulates the capacity offload problem as a non-cooperative game and sets some basic assumptions for analysis. Section \ref{sect:Nash equilibria of the model} establishes the existence of pure-strategy NE and the non-existence of mixed-strategy NE for COGs, and characterizes the structure of the NEs in two-player COGs. Section \ref{sect:convergence} analyzes the convergence properties of best-response learning algorithms for approaching NEs. Section \ref{sect:simulation} provides numerical results to illustrate the behaviors of learning algorithms. Finally, Section \ref{sect:conclusion} concludes the paper.

\section{\textsc{Model and Game-theoretic Considerations}}
\label{sect:models}

\subsection{System model}
\label{sect:system model}

\begin{figure}[!ht]
\centering
\subfigure[Transmission in private bands]{
\label{fig:mathmodel_private}
\includegraphics[width=3in]{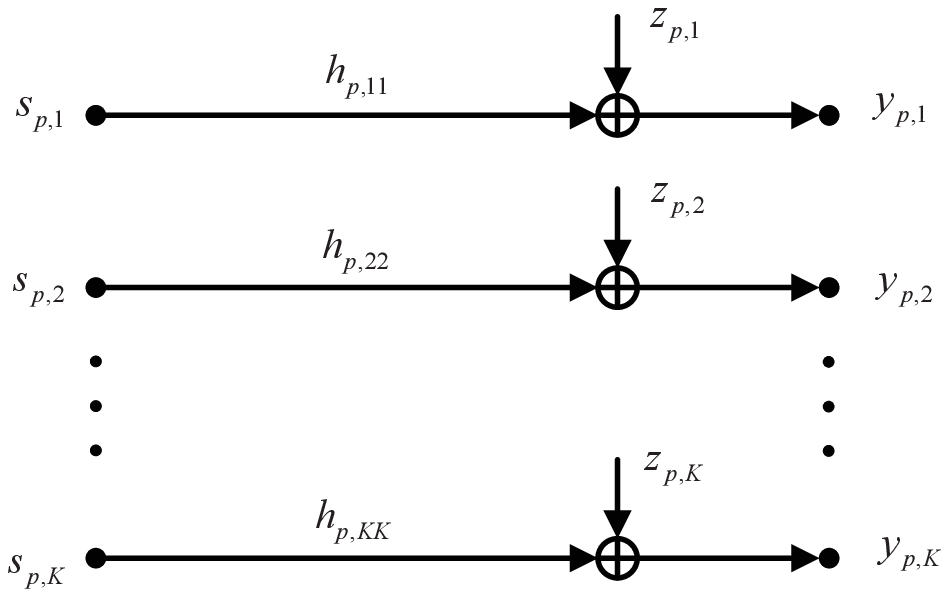}}
\subfigure[Transmission in the shared band]{
\label{fig:mathmodel_sharing}
\includegraphics[width=3in]{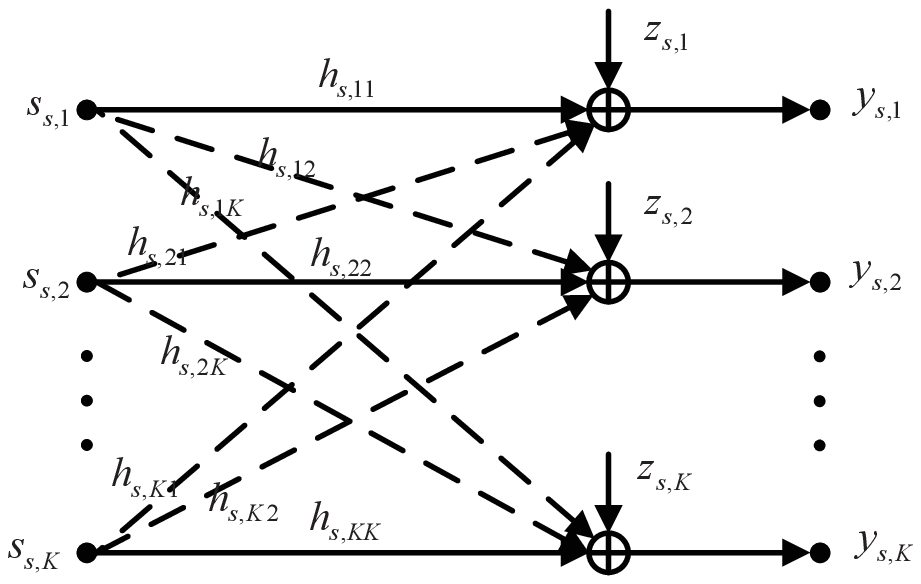}
}
\caption{Mathematical model of capacity offload}
\label{fig:mathmodels}
\end{figure}

An abstract model for capacity offloading may be described as follows; also see Figure \ref{fig:mathmodels}. Assume that a set, $\mathcal{K}=\{1,\cdots,K\}$, of access points from different WSPs are deployed in a geographic area. Suppose that each WSP, say WSP $k$, occupies a private band, whose bandwidth is $B_{p,k}$ Hertz. A shared band, whose bandwidth is $B_s$ Hertz, can be accessed and made use of by all WSPs. So the total available bandwidth, $B$ Hertz, consists of all WSP's private bands and the shared band; that is, $B = (\alpha + \sum_{k\in\mathcal{K}}\beta_{k}) B$ Hertz, where $\alpha={B_s}/{B}$ and $\beta_k={B_{p,k}}/{B}$, for $k \in \mathcal{K}$.

We assume that the transmitter of WSP $k$ has an average power constraint $P_{k}$, and that it can arbitrarily allocate its power on its own private band and the shared band. Denote by $x_{k}\in[0,1]$ the fraction of WSP $k$'s power allocated on its private band; that is, the WSP transmits at power $x_{k}P_{k}$ in its private band, and at $(1-x_{k})P_{k}$ in the shared band.\footnote{It is reasonable for each WSP to use up all its power. If there is power left unallocated for an WSP, it can allocate this residual power on its private band to increase its achievable rate without creating extra interference to other WSPs.} WSPs deploy time-division channel access scheme, so that in each time slot, a WSP exclusively transmits information to a single user equipment. All receivers are subject to additive white Gaussian noise with zero mean and power spectral density $n_{0}$. We further assume that all the link channels are frequency-flat, among which $|h_{k,k}|^2$ is the channel gain of the link from WSP $k$'s transmitter to its corresponding receiver in both its private band and the shared band, and $|h_{j,k}|^2$ is the channel gain of the link from WSP $j$'s transmitter to WSP $k$'s receiver.

The considered frequency-flat fading model is somewhat restrictive in that the channel gains over the private bands and the share band are set identical. This would occur mainly in situations where the private bands and the shared band are not far apart in frequency. Using such a simplified model, however, we are able to convey our key ideas effectively without dwelling in tedious technicalities. Extensions to more general fading models are possible; see, e.g., \cite{mochaourab2009resource}. In the sequel, we use $c_{j,k}$ to replace ${|h_{j,k}|^{2}}/{B}$ and $c_{k,k}$ to replace ${|h_{k,k}|^{2}}/{B}$, to simplify notations.

We denote the symbols transmitted by WSP $k$'s transmitter by $s_{p,k}$ (in its private band) and $s_{s,k}$ (in the shared band), with time indices suppressed. So the received baseband signals at WSP $k$'s receiver can be written as $y_{p,k}=h_{k,k}s_{p,k}+z_{p,k}$ (in its private band) and $y_{s,k}=h_{k,k}s_{s,k}+\sum_{j\in\mathcal{K}\backslash\{k\}}h_{j,k}s_{s,j}+z_{s,k}$ (in the shared band), where $\mathbb{E}[|z_{p,k}|^{2}]=n_{0}\beta_{i} B$, $\mathbb{E}[|z_{s,k}|^{2}]=n_{0}\alpha B$, $\mathbb{E}[|s_{p,k}|^{2}]=x_{k}P_{k}$ and $\mathbb{E}[|s_{s,k}|^{2}]=(1-x_{k})P_{k}$. We consider a naive coding scheme in which all transmitted symbols follow Gaussian codebooks and all receivers adopt single-user decoding treating others' signals as noise. Hence, WSP $k$'s achievable rate (normalized by $B$) is:
\begin{equation}
\label{equ:SumRate}
\begin{aligned}
u_{k}(\mathbf{x})=
& \underbrace{\alpha \log_2 \left(1+\frac{(1-x_{k})P_{k}c_{k,k}}{n_{0}\alpha+\sum_{j\in\mathcal{K}\backslash\{k\}}(1-x_{j})P_{j}c_{j,k}}\right)}_{\mathrm{in\; the\; shared\; band}}\\
& \underbrace{+\beta_{k} \log_2 \left(1+\frac{x_{k}P_{k}c_{k,k}}{n_{0}\beta_{k}}\right) \quad (\mathrm{bits} / \mathrm{Hertz})}_{\mathrm{in\; the\; private\; band}},
\end{aligned}
\end{equation}
where vector $\mathbf{x}=(x_{1},\cdots,x_{K})^{T}$ represents the power allocation strategies adopted by all the WSPs. Due to the lack of coordination among WSPs, it is reasonable to suppose that each WSP adopts its individual strategy to attempt to maximize its own achievable rate, regardless of other WSPs' rates. This situation is formally described by the COG, introduced in the following.

\subsection{Game-theoretic model}
\label{sect:Game-Theoretic Model}
We define the pure-strategy form of COG as $\mathcal{G}=\langle \mathcal{K},(X_{k})_{k\in \mathcal{K}}, (u_{k})_{k\in \mathcal{K}}\rangle$, in which $\mathcal{K}$ represents the set of players (i.e., the WSPs), $X_{k}$ is the set of WSP $k$'s pure strategies, i.e., the power allocation ratio $x_k \in [0, 1]$, and the utility function $u_{k}$ is given by (\ref{equ:SumRate}). More generally, we may also consider mixed strategies: each player $k\in \mathcal{K}$ chooses its strategy $x_{k}$ following a discrete probability distribution $\mathbf{\pi}_k(x_k)=(\pi_{k,x_{k}^{(1)}},\cdots,\pi_{k,x_{k}^{(N)}})^{T}\in \Delta(X_{k})$ over a finite strategy space (where $\pi_{k,x_{k}^{(n)}}$ represents the probability that WSP $k$ chooses strategy $x_{k}^{(n)}$) or following a probability distribution function $\pi_{k}(x_{k})\in \Delta (X_{k})$ over a continuous strategy space, where $\int^{1}_{0}\pi_{k}(x_{k})\mathrm{d}x_{k}=1$, $\pi_{k}(x_{k})\geq 0$. For mixed strategies, the utility function $\bar{u}_{k}$ of player $k$ is defined as the expectation of $u_{k}$ in (\ref{equ:SumRate}) with respect to the probability distributions of all players' strategies. We can define the mixed strategic-form game for COG as $\mathcal{G}^{'}=\langle \mathcal{K},(\Delta (X_{k}))_{k\in \mathcal{K}}, (\bar{u}_{k})_{k\in \mathcal{K}}\rangle$.

The definitions of pure-strategy and mixed-strategy NEs are as follows \cite{fudenberg1991game}.
\begin{defn}[Pure-strategy Nash equilibrium]
\label{defn:PNE}
A strategy profile $\mathbf{x}^{*}=(x_{1}^{*},...,x_{K}^{*})^{T}$ is a pure-strategy NE for the COG, if $\forall k\in \mathcal{K}$ and $\forall x_{k}^{'}\in X_{k}$, $u_{k}(x_{k}^{*},\mathbf{x}_{-k}^{*})\geq u_{k}(x_{k}^{'},\mathbf{x}_{-k}^{*})$, where the subscript $-k$ represents all the players other than player $k$.
\end{defn}
\begin{defn}[Mixed-strategy Nash equilibrium]
\label{defn:MNE}
A mixed-strategy profile $\mathbf{\pi^{*}}$ is a mixed-strategy NE for the COG, if $\forall k\in \mathcal{K}$ and $\forall \pi_{k}^{'}\in \Delta (X_{k})$, $\bar{u}_{k}(\pi_{k}^{*},\mathbf{\pi}_{-k}^{*})\geq \bar{u}_{k}(\pi_{k}^{'},\mathbf{\pi}_{-k}^{*})$, where $\bar{u}_{k}(\mathbf{\pi})=\mathbb{E}_{\pi}[u_{k}(x_{k},\mathbf{x}_{-k})]: \Delta (X_{1})\times...\times \Delta (X_{K})\longrightarrow\mathbb{R}$.
\end{defn}

\section{\textsc{Analysis of Nash Equilibria}}
\label{sect:Nash equilibria of the model}

In this section, we establish the existence of pure-strategy NE, as well as the nonexistence of mixed-strategy NE, for COGs. With the aid of explicit best-response function, we fully characterize the structure of NEs for two-player COGs, by displaying the relationship between network parameters and the NEs.

\subsection{Existence of Pure-strategy NE}
\label{sect:NE existence}

The existence of pure-strategy NEs for COGs is an application of the following lemma.
\begin{lem}[Rosen \cite{rosen1965existence}]
\label{lem:concave N person}
At least one NE exists for every concave $K$-player game.
\end{lem}

For a concave game, the joint strategy set is convex, closed and bounded, and for each player $k \in \mathcal{K}$, its utility function is continuous with respect to the joint strategy and is concave with respect to player $k$'s strategy. Then we state the existence theorem as follows:
\begin{thm}[Existence of pure-strategy NE]
\label{thm:NE existence}
A pure-strategy COG, $\mathcal{G}=\langle \mathcal{K},(X_{k})_{k\in \mathcal{K}}, (u_{k})_{k\in \mathcal{K}}\rangle$, always possesses at least one NE, i.e., there is at least one strategy profile $\mathbf{x}^{*}=(x_{1}^{*},\cdots,x_{K}^{*})^{T}$ such that $u_{k}(x_{k}^{*},\mathbf{x}_{-k}^{*})\geq u_{k}(x_{k}^{'},\mathbf{x}_{-k}^{*})$ holds, $\forall k\in \mathcal{K}$, $\forall x_{k}^{'}\in X_{k}$.
\end{thm}
\begin{IEEEproof}
By lemma \ref{lem:concave N person}, it suffices to verify that the pure-strategy COG is a concave $K$-player game:
\begin{enumerate}
\item The pure-strategy $\mathbf{x}$ is from a convex, closed and bounded set, namely $X_{1}\times \cdots \times X_{K}=\{ \mathbf{x}=(x_1,\cdots,x_K) | x_{k} \in [0,1], \mathrm{for} \; k=1,\cdots, K\}$;
\item Player $k$'s utility function, $u_{k}(\mathbf{x})$, defined by (\ref{equ:SumRate}), is continuous in $\mathbf{x}$ and is concave in $x_{k}$, since
    \begin{equation}
    \begin{aligned}
    \label{equ:concave}
    \frac{\partial^{2} u_{k}(\mathbf{x})}{\partial x_{k}^{2}}=
     -\frac{\alpha P_{k}^{2}c_{k,k}^{2}}{[n_{0}\alpha +\sum_{j\in \mathcal{K}}(1-x_{j})P_{j}c_{j,k}]^{2}} -\frac{\beta_{k}P_{k}^{2}c_{k,k}^{2}}{(n_{0}\beta_{k}+x_{k}P_{k}c_{k,k})^{2}} < 0.\nonumber
    \end{aligned}
    \end{equation}
\end{enumerate}
\end{IEEEproof}

\subsection{Nonexistence of Mixed-strategy NE}
\label{sect:mixed nonexistence}

Theorem \ref{thm:NE existence} guarantees that at least one pure-strategy NE exists, but it is remains unclear whether a mixed-strategy NE exists. In the following, we rule out this possibility.
\begin{thm}[Nonexistence of mixed-strategy NE]
\label{thm:mixed NE nonexistence}
For a COG, there is no mixed strategy that makes the players reach an NE.\footnote{Pure strategies are degenerated mixed strategies, and are excluded from the consideration.}
\end{thm}
\begin{IEEEproof}
For a given $k$, let us fix players $-k$'s strategy probability distribution as $\pi_{-k}^{*}(\mathbf{x}_{-k})\in \Delta(X_{-k})$. Regarding the utility $\bar{u}_{k}$ of player $k$, we have
\begin{equation}
\label{equ:jensen's inequality}
\begin{aligned}
\bar{u}_{k}(\pi_{k},\pi_{-k}^{*})
& =\int_{X_{-k}}\!\!\int_0^1 \!\!u_k(x_k,\mathbf{x}_{-k})\pi_k(x_k)\pi_{-k}^{*}(\mathbf{x}_{-k})\mathrm{d}x_k\mathrm{d}\mathbf{x}_{-k} \nonumber \\
& =\int_{X_{-k}}\!\!\int_0^1 \!\! \bigg[\alpha \log_2 \left(1+\frac{(1-x_{k})P_{k}c_{k,k}}{n_{0}\alpha+\sum_{j\in \mathcal{K}\backslash \{k\}}(1-x_{j})P_{j}c_{j,k}} \right) \nonumber \\
& \quad +\beta_{k}\log_2 \left(1+\frac{x_{k}P_{k}c_{k,k}}{n_{0}\beta_{k}}\right) \bigg] \pi_k(x_k)\pi_{-k}^{*}(\mathbf{x}_{-k})\mathrm{d}x_k\mathrm{d}\mathbf{x}_{-k} \nonumber \\
& \stackrel{(\mathrm{a})}{\leq} \int_{X_{-k}}\!\! \bigg[ \alpha\log_2 \left(1+\frac{\int_0^1\!(1-x_{k})P_{k}c_{k,k}\pi_k(x_k)\mathrm{d}x_k}{n_{0}\alpha+\sum_{j\in \mathcal{K}\backslash \{k\}}(1-x_{j})P_{j}c_{j,k}}\right) \nonumber \\
& \quad +\beta_{k}\log_2 \left(1+\frac{\int_0^1\!x_{k}P_{k}c_{k,k}\pi_k(x_k)\mathrm{d}x_k}{n_{0}\beta_{k}}\right)\bigg] \pi_{-k}^{*}(\mathbf{x}_{-k})\mathrm{d}\mathbf{x}_{-k} \nonumber \\
& =\int_{X_{-k}}\!\! \bigg[\alpha \log_2 \left(1+\frac{(1-\mathbb{E}[x_{k}])P_{k}c_{k,k}}{n_{0}\alpha+\sum_{j\in \mathcal{K}\backslash \{k\}}(1-x_{j})P_{j}c_{j,k}}\right) \nonumber \\
& \quad +\beta_{k}\log_2 \left(1+\frac{\mathbb{E}[x_{k}]P_{k}c_{k,k}}{n_{0}\beta_{k}}\right)\bigg]\pi_{-k}^{*}(\mathbf{x}_{-k})\mathrm{d}\mathbf{x}_{-k},
\end{aligned}
\end{equation}
where (a) follows from Jensen's inequality. Since $u_k(\mathbf{x})$ is strictly concave in $x_k$ (cf. proof of Theorem \ref{thm:NE existence}), we achieve equality in (a) if and only if $x_k=\mathbb{E}[x_k]$ with probability $1$, which means that player $k$'s strategy set is deterministic. So it is optimal for player $k$ to adopt a pure strategy to maximize its utility, and the same conclusion also applies to every other player. Thus we rule out the possibility for mixed strategies being optimal and complete the proof.
\end{IEEEproof}

According to Theorem \ref{thm:mixed NE nonexistence}, in the following we only need to focus on pure strategies.

\subsection{Best-response Functions}

Since $u_k(\mathbf{x})$ is strictly concave in $x_k$, player $k$ may choose a pure strategy $x_k\in [0,1]$ satisfying
\begin{equation}
\label{equ:solve maximum}
\frac{\partial u_k(x_k,\mathbf{x}_{-k})}{\partial x_k}=0,
\end{equation}
to maximize its utility when its opponent players adopt strategy $\mathbf{x}_{-k}$. Solving (\ref{equ:solve maximum}), we get
\begin{equation}
\label{equ:linear response}
x_k \triangleq \mathrm{f}_k(\mathbf{x}_{-k}) = \frac{\beta_k}{\alpha+\beta_k}\left[1+\frac{\sum_{j\in \mathcal{K}\backslash \{k\}}P_{j}c_{j,k}(1-x_j)}{P_k c_{k,k}}\right].
\end{equation}
Taking into account that $x_k$ is no greater than one, we obtain the best-response function $\mathrm{BR}_k$ of player $k$ as
\begin{eqnarray}
\label{equ:Kuser best response function}
x_k &\triangleq& \mathrm{BR}_k(\mathbf{x}_{-k})\nonumber\\
&=& \mathrm{min}\left\{\frac{\beta_k}{\alpha+\beta_k}\left[1+\frac{\sum_{j\in \mathcal{K}\backslash \{k\}}P_{j}c_{j,k}(1-x_j)}{P_k c_{k,k}}\right],1\right\}.
\end{eqnarray}
We can determine the NEs of a COG by solving the equations
\begin{eqnarray}
x_k = \mathrm{BR}_k(\mathbf{x}_{-k}),\quad\mbox{for}\; k=1,\cdots,K.
\end{eqnarray}
Note that the equations may have multiple solutions, corresponding to multiple NEs, as illustrated in the next subsection.

\subsection{Characterization of NEs for Two-player COGs}
\label{sect:2playerCOG}

When a COG has only two players, it is convenient to describe the relationship between the number and behavior of NEs and network parameters. The two players' best-response functions are:
\begin{equation}
\label{equ:2bestresponse1}
x_1 = \mathrm{BR}_1(x_2) = \mathrm{min}\left\{\frac{\beta_1}{\alpha+\beta_1}\left[1+\frac{P_2c_{2,1}}{P_1c_{1,1}}(1-x_2)\right],1\right\},
\end{equation}
\begin{equation}
\label{equ:2bestresponse2}
x_2 = \mathrm{BR}_2(x_1) = \mathrm{min}\left\{\frac{\beta_2}{\alpha+\beta_2}\left[1+\frac{P_1c_{1,2}}{P_2c_{2,2}}(1-x_1)\right],1\right\}.
\end{equation}

\begin{figure}[!ht]
\centering
\includegraphics[width=3in,
height=3in]{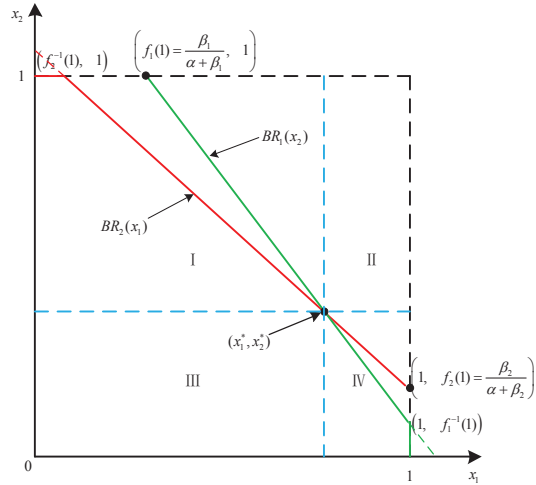}
\caption{Case 1) of Theorem \ref{thm:2-player-NE}. The network parameters satisfy $f_2^{-1}(1)<f_1(1)$ and $f_1^{-1}(1)<f_2(1)$, and thus the two best-response functions intersect at only one point as indicated by $(x_1^\ast, x_2^\ast)$, corresponding to the unique NE in the COG.}
\label{fig1}
\end{figure}

\begin{figure}[!ht]
\centering
\includegraphics[width=3in,
height=3in]{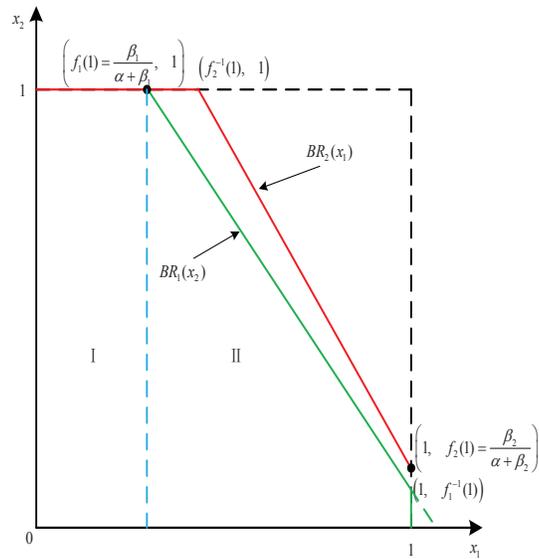}
\caption{Cases 2) and 3) of Theorem \ref{thm:2-player-NE}. Here we depict the situation for case 2) only, and that for case 3) is similar. The network parameters satisfy $f_2^{-1}(1) \geq f_1(1)$ and $f_1^{-1}(1)<f_2(1)$, and thus the two best-response functions intersect only at $\left(f_1(1) = \frac{\beta_1}{\alpha + \beta_1}, 1\right)$.}
\centering
\label{fig2}
\end{figure}

\begin{figure}[!ht]
\centering
\includegraphics[width=3in,
height=3in]{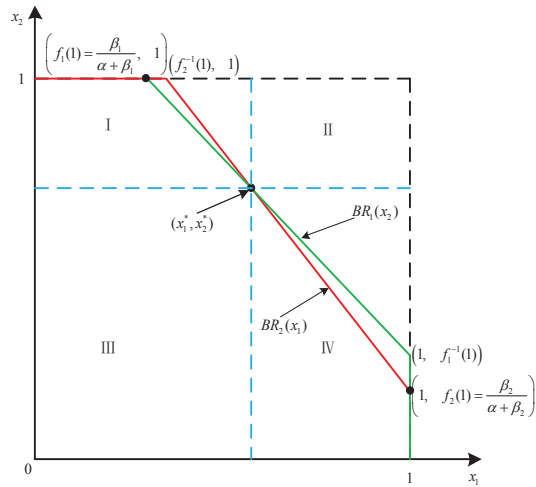}
\caption{Case 4) of Theorem \ref{thm:2-player-NE}. The network parameters satisfy $f_2^{-1}(1)> f_1(1)$ and $f_1^{-1}(1)>f_2(1)$, and thus the two best-response functions intersect at three points, as indicated in the figure.}
\label{fig4}
\end{figure}

\begin{figure}[!ht]
\centering
\includegraphics[width=3in,
height=3in]{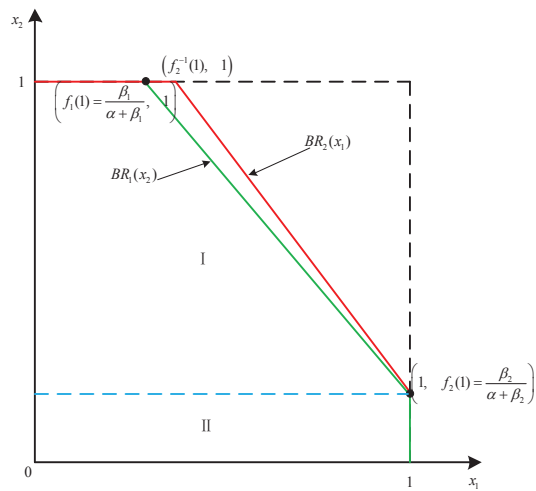}
\caption{The special subcase with two NEs of case 4). The network parameters satisfy $f_2^{-1}(1)> f_1(1)$ and $f_1^{-1}(1)=f_2(1)$, and thus the intersecting point $(x_1^\ast, x_2^\ast)$ in Figure \ref{fig4} coincides with $\left(1, f_2(1) = \frac{\beta_2}{\alpha + \beta_2}\right)$. The other possibility $f_2^{-1}(1)= f_1(1)$ and $f_1^{-1}(1)>f_2(1)$ is similar and thus omitted for conciseness.}
\label{fig3}
\end{figure}

\begin{figure}[!ht]
\centering
\includegraphics[width=3in,
height=3in]{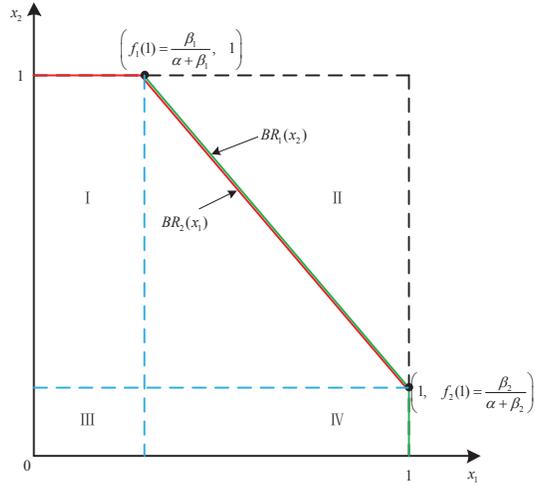}
\caption{The special subcase with an infinite number of NEs of case 4). The network parameters satisfy $f_2^{-1}(1)= f_1(1)$ and $f_1^{-1}(1)=f_2(1)$, and thus the two slope segments of the best-response functions completely coincide, leading to an infinite number of NEs.}
\label{fig5}
\end{figure}

When we draw both players' best-response functions in the same $(x_1, x_2)$-coordinate diagram, by the definition of NE, the intersection points of the two best-response functions correspond to the NEs of the COG. Depending upon the network parameters, the number and locations of the NEs vary, as illustrated in Figures \ref{fig1}-\ref{fig5}. The behavior of NEs can be summarized in the following theorem.
\begin{thm}
\label{thm:2-player-NE}
For a two-player COG, it has
\begin{enumerate}
\item a unique NE as the solution of
\begin{eqnarray}
x_1 &=& \frac{\beta_1}{\alpha+\beta_1}\left[1+\frac{P_2c_{2,1}}{P_1c_{1,1}}(1-x_2)\right],\nonumber\\
x_2 &=& \frac{\beta_2}{\alpha+\beta_2}\left[1+\frac{P_1c_{1,2}}{P_2c_{2,2}}(1-x_1)\right],\nonumber
\end{eqnarray}
when the network parameters satisfy (see Figure \ref{fig1})
\begin{eqnarray}
\label{eqn:case-1}
\frac{c_{1,2}}{c_{2,2}} < \frac{P_2}{P_1} \frac{\alpha+\beta_1}{\beta_2}
       \quad \mathrm{and} \quad
       \frac{c_{2,1}}{c_{1,1}}<\frac{P_1}{P_2}\frac{\alpha+\beta_2}{\beta_1};
\end{eqnarray}
\item a unique NE as $(x_1, x_2) = \left(\frac{\beta_1}{\alpha + \beta_1}, 1\right)$, when (see Figure \ref{fig2})
\begin{eqnarray}
\frac{c_{1,2}}{c_{2,2}} \geq \frac{P_2}{P_1} \frac{\alpha+\beta_1}{\beta_2}
       \quad \mathrm{and} \quad
       \frac{c_{2,1}}{c_{1,1}} < \frac{P_1}{P_2}\frac{\alpha+\beta_2}{\beta_1};
\end{eqnarray}
\item a unique NE as $(x_1, x_2) = \left(1, \frac{\beta_2}{\alpha + \beta_2}\right)$, when (see Figure \ref{fig2})
\begin{eqnarray}
\frac{c_{1,2}}{c_{2,2}} < \frac{P_2}{P_1} \frac{\alpha+\beta_1}{\beta_2}
       \quad \mathrm{and} \quad
       \frac{c_{2,1}}{c_{1,1}} \geq \frac{P_1}{P_2}\frac{\alpha+\beta_2}{\beta_1};
\end{eqnarray}
\item three NEs as those listed in cases 1)-3) above together, when (see Figure \ref{fig4})
\begin{eqnarray}
       \label{equ:2user 3NE}
       \frac{c_{1,2}}{c_{2,2}} > \frac{P_2}{P_1} \frac{\alpha+\beta_1}{\beta_2}
       \quad \mathrm{and} \quad
       \frac{c_{2,1}}{c_{1,1}} > \frac{P_1}{P_2}\frac{\alpha+\beta_2}{\beta_1}.
\end{eqnarray}
For this case, there are two singular subcases. If any one of the two inequalities becomes equality, the three NEs collapse into two, since the NE in case 1) coincides with the NE in case 2) or 3); see Figure \ref{fig3}. If both of the two inequalities become equal, there are an infinite number of NEs since the two lines
\begin{eqnarray}
x_1 &=& \frac{\beta_1}{\alpha+\beta_1}\left[1+\frac{P_2c_{2,1}}{P_1c_{1,1}}(1-x_2)\right],\nonumber\\
x_2 &=& \frac{\beta_2}{\alpha+\beta_2}\left[1+\frac{P_1c_{1,2}}{P_2c_{2,2}}(1-x_1)\right]\nonumber
\end{eqnarray}
coincide for all $x_1 \in \left[\frac{\beta_2}{\alpha + \beta_2}, 1\right]$; see Figure \ref{fig5}.
\end{enumerate}
\end{thm}

Let us make a few comments regarding Theorem \ref{thm:2-player-NE}. Throughout its cases, we notice that the behavior of NEs depends upon the comparison between two sets of quantities,
\begin{eqnarray}
\frac{c_{1, 2}P_1}{\alpha + \beta_1} \;\mbox{versus}\; \frac{c_{2, 2}P_2}{\beta_2}, \quad\mbox{and}\quad \frac{c_{2, 1}P_2}{\alpha + \beta_2} \;\mbox{versus}\; \frac{c_{1, 1}P_1}{\beta_1}.
\end{eqnarray}
In the first comparison, $c_{1, 2}P_1/(\alpha + \beta_1)$ can be viewed as the interference strength to player $2$, when player $1$ evenly distributes its power across its private band and the shared band, and $c_{2, 2} P_2/\beta_2$ can be viewed as the signal strength to player $2$ when it exclusively uses its private band. The two quantities in the second comparison can also be interpreted correspondingly. It is interesting that the comparison between these seemingly unrelated quantities determines the behavior of NEs in the game.

The NE in case 1) implies that both players allocate a portion of their power budgets to the shared band, so that they indeed coexist tolerating a certain amount of interference. According to (\ref{eqn:case-1}), this occurs when the interference strengths to both players are weak. The NE in case 2) or 3), $\left(\frac{\beta_1}{\alpha + \beta_1}, 1\right)$ or $\left(1, \frac{\beta_2}{\alpha + \beta_2}\right)$, implies that player $2$ or player $1$ completely retreats from the shared band, while the other player evenly distributes its power across its private band and the shared band. This is the unique stable operating point (i.e., NE) when one player experiences strong interference while the other's interference is still weak. In case 4), the interference strengths to both players are strong, and it is interesting that then the system may reach any of three equilibrium operating points: players coexisting within the shared band, and either player completely retreating from the shared band.

\section{\textsc{Best-response Dynamics for Distributed Learning}}
\label{sect:convergence}

Having established the existence of NEs in COGs, in this section, we focus on the behavior of best-response dynamics for distributed learning, to examine whether the dynamics' evolution eventually leads all players to reach an NE. We begin with defining the best-response dynamic learning procedures. We then fully characterize the convergence properties of best-response dynamics for two-player COGs. Finally for $K$-player COGs, we establish the convergence properties of best-response dynamics, when the network parameters are such that the resulting best-response functions are all linear.

\subsection{Best-response dynamics}

We assume that in a best-response dynamic process, whenever a player $k$ decides to update its strategy at time instant $t$, it possesses the knowledge of the other players' joint strategies $\mathbf{x}_{-k}$ at $t$, and the updating rule is
\begin{equation}
x_k(t^+)=\mathrm{BR}_k(\mathbf{x}_{-k}(t)).
\end{equation}
That is, the updated strategy $x_k$ right after the time instant $t$ is the one that maximizes the utility $u_k$ given the strategies of the other players at $t$. Inspecting the best-response functions (\ref{equ:Kuser best response function}), we note that in order to update $x_k$, it is only necessary for WSP $k$'s receiver to measure the level of aggregated interference in the shared band, $\sum_{j \in \mathcal{K}\backslash\{k\}} P_j c_{j, k} (1 - x_j)$. Furthermore, we may rewrite (\ref{equ:Kuser best response function}) as
\begin{eqnarray}
x_k = \mathrm{min}\left\{\frac{\beta_k}{\alpha+\beta_k}\left[1+\frac{1 - x_k}{\mathrm{SIR}_k}\right],1\right\},
\end{eqnarray}
where
\begin{eqnarray}
\mathrm{SIR}_k \triangleq \frac{P_k c_{k, k} (1 - x_k)}{\sum_{j\in \mathcal{K}\backslash \{k\}}P_{j}c_{j,k}(1-x_j)}
\end{eqnarray}
denotes the signal-to-interference ratio (SIR) of WSP $k$'s receiver in the shared band.

In the following, we consider two situations. In the first situation, all the players update their strategies simultaneously. We call this simultaneous-move best-response dynamic (SMBRD). In the second situation, the players update their strategies sequentially, in a periodic round robin fashion. We call this alternating-move best-response dynamic (AMBRD).

For the two kinds of best-response dynamics, we propose the following distributed learning algorithms. In the algorithm description, we assume for simplicity that time is slotted and updates occur at the beginning of each time slot.

{\bf Simultaneous-Move Best-Response Dynamic (SMBRD):}
\newline\noindent\rule{1\textwidth}{0.5pt}

\noindent For each player $k \in \mathcal{K}$:
\begin{description}
\item[Step 1:]
\quad At time $t=0$, player $k$ selects an initial strategy $x_k(0)$ arbitrarily within $[0, 1]$;
\item[Step 2:]
\quad At time $t+1$, given the measurement of $\mathrm{SIR}_k(t)$, player $k$ updates its strategy to $x_k(t+1)= \mathrm{min}\left\{\frac{\beta_k}{\alpha+\beta_k}\left[1+\frac{1}{\mathrm{SIR}_k(t)}\right],1\right\}$;
\item[Step 3:]
\quad Increase the time slot index to $t+1$ and go back to step 2 until all the players' strategies become stationary or the time index reaches the prescribed maximum number of iterations.
\end{description}

\noindent\rule{1\textwidth}{0.5pt}\\[-.5cm]
\noindent\rule[0.15\baselineskip]{1\textwidth}{0.5pt}

{\bf Alternating-Move Best-Response Dynamic (AMBRD):}
\newline\noindent\rule{1\textwidth}{0.5pt}

\noindent Without loss of generality, we assume that the $K$ players update their strategies sequentially and periodically, from player $1$ until player $K$ in each updating cycle.

\begin{description}
\item[Step 1:]
\quad At time $t=0$, player $1$ selects an initial strategy $x_1(0)$ arbitrarily within $[0, 1]$;
\item[Step 2:]
\quad For each player $k=2,\cdots,K$, it takes turns to revise its strategy at time $t+k-1$. Given the measurement of $\mathrm{SIR}_k(t+k-1)$, player $k$ updates its strategy to $x_k(t+k)= \mathrm{min}\left\{\frac{\beta_k}{\alpha+\beta_k}\left[1+\frac{1}{\mathrm{SIR}_k(t+k-1)}\right],1\right\}$;
\item[Step 3:]
\quad Increase the time index to $t+K$, and player $1$ updates its strategy according to $x_1(t+K)= \mathrm{min}\left\{\frac{\beta_1}{\alpha+\beta_1}\left[1+\frac{1}{\mathrm{SIR}_1(t+K-1)}\right],1\right\}$. Go back to step 2 until all the players' strategies become stationary or the time index reaches the prescribed maximum number of iterations.
\end{description}

\noindent\rule{1\textwidth}{0.5pt}\\[-.5cm]
\noindent\rule[0.15\baselineskip]{1\textwidth}{0.5pt}

\subsection{Convergence Analysis for Two-player COGs}

In this subsection, we analyze the simple case where there are only two players in the COG. Let us begin with the SMBRD, whose execution exhibits a sequence of strategies of the two players. From the best-response functions of the two players, we see that the sequence of strategies can be decomposed into the following two subsequences
\begin{eqnarray}
\label{equ:sequences}
&&\mathrm{Sequence} [\mathrm{a}]: \left(x_1(0),x_2(1)\right),\left(x_1(2),x_2(3)\right),\cdots,\nonumber\\
\; \mathrm{and} \;
&&\mathrm{Sequence} [\mathrm{b}]: \left(x_2(0),x_1(1)\right),\left(x_2(2),x_1(3)\right),\cdots,\nonumber
\end{eqnarray}
each of which evolves independently of the other. Sequence[a] is uniquely determined by player $1$'s initial strategy $x_1(0)$, and Sequence[b] by $x_2(0)$. So we can study each sequence's convergence property individually. Only when these two subsequences' limits coincide, the SMBRD converges; otherwise the SMBRD exhibits a cycling behavior between the limits of the two subsequences. In the following discussion, for notational simplicity, we abbreviate (\ref{equ:linear response}) as $x_k = f_k(x_{-k})=-b_k x_{-k}+a_k$, where $a_k,b_k>0$ for $k = 1, 2$, and denote the intersection of the two lines by $(x_1^{*},x_2^{*})$. We then discuss the convergence property of the two sequences under difference network parameters.

\begin{enumerate}
\item When the COG only has a unique NE:
    \begin{enumerate}
    \item When condition:
    \begin{equation}
    \label{equ:2user 1NE condition11}
    \frac{c_{1,2}}{c_{2,2}} < \frac{P_2}{P_1} \frac{\alpha+\beta_1}{\beta_2}\;\mathrm{and}\;
    \frac{c_{2,1}}{c_{1,1}}<\frac{P_1}{P_2}\frac{\alpha+\beta_2}{\beta_1}
    \end{equation}
    is satisfied, the unique NE is $(x_1^\ast, x_2^\ast)$, the solution of $x_k = -b_k x_{-k} + a_k$ ($k = 1, 2$); see Figure \ref{fig1}. By symmetry we only discuss Sequence[a]'s convergence property and the discussion for Sequence[b] is similar. From the conditions (\ref{equ:2user 1NE condition11}), if $x_1(t)>f_2^{-1}(1)$, we have
    \begin{equation}
    \label{equ:2user convergence}
    x_1(t+2)-x_1^\ast = b_1b_2 [x_1(t)-x_1^\ast],
    \end{equation}
    in which $0 < b_1b_2 < 1$. So the updating process of $x_1(t)$ will converge to $x_1^\ast$. If $x_1(t)<f_2^{-1}(1)$, then we have $x_2(t+1) = 1$, $x_1(t+2)=f_1(1) > f_2^{-1}(1)$, which again enters the regime of (\ref{equ:2user convergence}) and thus $x_1(t)$ will converge to $x_1^\ast$ and $x_2(t)$ will converge to $\mathrm{BR}_2(x_1^\ast)=x_2^\ast$. The analysis of Sequence[b] yields the same result, and we therefore see that under (\ref{equ:2user 1NE condition11}), both Sequence[a] and Sequence[b] converge to the unique NE.
    \item when condition:
    \begin{equation}
    \label{equ:2user 1NE condition2}
    \frac{c_{1,2}}{c_{2,2}} \geq \frac{P_2}{P_1} \frac{\alpha+\beta_1}{\beta_2}\;\mathrm{and}\;
    \frac{c_{2,1}}{c_{1,1}}<\frac{P_1}{P_2}\frac{\alpha+\beta_2}{\beta_1}
    \end{equation}
    is satisfied (cf. Figure \ref{fig2}), similar to the discussion above, we can verify that both Sequence[a] and Sequence[b] converge to the NE $(f_1(1),1)$ given any initial strategies $(x_1(0),x_2(0))$. For the other symmetric case, both sequences will converge to the NE $(1,f_2(1))$.
    \end{enumerate}
    Therefore, when the COG has only one NE, SMBRD will always converge to the NE.

\item When the COG has two NEs, by symmetry we only discuss the case where conditions
\begin{equation}
\label{equ:2user 2NE condition}
\frac{c_{1,2}}{c_{2,2}} > \frac{P_2}{P_1} \frac{\alpha+\beta_1}{\beta_2}\;\mathrm{and}\;
\frac{c_{2,1}}{c_{1,1}}=\frac{P_1}{P_2}\frac{\alpha+\beta_2}{\beta_1}
\end{equation}
are satisfied (cf. Figure \ref{fig3}). Note that with the conditions, $b_1b_2 > 1$. Similar to the discussion above, we see that when $x_1(0) < 1$, Sequence[a] converges to $(f_1(1),1)$, but when $x_1(0)=1$, Sequence[a] converges to $(1,f_2(1))$. For Sequence[b], when $x_2(0) \leq f_2(1)$, we have $x_1(1)=1$ and $x_2(2)=f_2(1)$, and then Sequence[b] converges to $(1,f_2(1))$; when $x_2(0) > f_2(1)$, due to $b_1b_2>1$ and
\begin{equation}
\label{equ:2user convergenceb}
x_2(t+2)-x_2^\ast= b_1b_2 [x_2(t)-x_2^\ast],
\end{equation}
Sequence[b] converges to $(f_1(1),1)$. Therefore, under the conditions (\ref{equ:2user 2NE condition}), when the initial strategies $\mathbf{x}(0)$ fall within region I in Figure \ref{fig3}, SMBRD converges to the NE $(f_1(1),1)$, and when the initial strategies $\mathbf{x}(0)$ fall within region II, SMBRD ends up with cycling between strategies $(f_1(1), f_2(1))$ and $(1, 1)$.

\item When the COG has three NEs, the conditions
\begin{equation}
\label{equ:2user 3NE condition}
\frac{c_{1,2}}{c_{2,2}} > \frac{P_2}{P_1} \frac{\alpha+\beta_1}{\beta_2}\;\mathrm{and}\;
\frac{c_{2,1}}{c_{1,1}}>\frac{P_1}{P_2}\frac{\alpha+\beta_2}{\beta_1}
\end{equation}
are satisfied (cf. Figure \ref{fig4}). We trace the evolution of the sequences and get the following result. For Sequence[a], when the initial strategy satisfies $x_1(0) < x_1^\ast$, it converges to the NE $(f_1(1),1)$, otherwise it converges to another NE $(1,f_2(1))$. For Sequence[b], when the initial strategy satisfies $x_2(0)<x_2^\ast$, it converges to the NE $(1,f_2(1))$, otherwise it converges to another NE $(f_1(1),1)$. Therefore, under the conditions (\ref{equ:2user 3NE condition}), when the initial strategies $\mathbf{x}(0)$ fall within region I in Figure \ref{fig4}, SMBRD converges to the NE $(f_1(1),1)$, when the initial strategies $\mathbf{x}(0)$ fall within region II or III, SMBRD ends up with cycling between strategies $(f_1(1),f_2(1))$ and $(1,1)$, and when the initial strategies $\mathbf{x}(0)$ fall within region IV, SMBRD converges to the NE $(1,f_2(1))$.

\item When the COG has an infinite number of NEs, the conditions
\begin{equation}
\label{equ:2user infNE condition}
\frac{c_{1,2}}{c_{2,2}} = \frac{P_2}{P_1} \frac{\alpha+\beta_1}{\beta_2}\;\mathrm{and}\;
\frac{c_{2,1}}{c_{1,1}}=\frac{P_1}{P_2}\frac{\alpha+\beta_2}{\beta_1}
\end{equation}
are satisfied (cf. Figure \ref{fig5}). For Sequence[a], when the initial strategy satisfies $x_1(0) < f_1(1)$, it converges to the NE $(f_1(1),1)$, otherwise it converges to the NE $(x_1(0),f_2(x_1(0)))$; for Sequence[b], when the initial strategy satisfies $x_2(0)<f_2(1)$, it converges to the NE $(f_1(x_2(0)),x_2(0))$, otherwise it converges to NE $(1,f_1(1))$. Therefore, under the conditions (\ref{equ:2user infNE condition}), when the initial strategies $\mathbf{x}(0)$ fall within region I in Figure \ref{fig5}, SMBRD ends up with cycling between strategies $(f_1(1),x_2(0))$ and $(f_1(x_2(0)),1)$, when $\mathbf{x}(0)$ falls within region II, SMBRD ends up with cycling between strategies $(f_1(x_2(0)),f_2(x_1(0)))$ and $(x_1(0),x_2(0))$, when $\mathbf{x}(0)$ falls within region III, SMBRD ends up with cycling between strategies $(f_1(1),f_2(1))$ and $(1,1)$, and when $\mathbf{x}(0)$ falls within region IV, SMBRD ends up with cycling between strategies $(x_1(0),f_2(1))$ and $(1,f_2(x_1(0)))$.
\end{enumerate}

Summarizing the discussion above for various conditions, we see that both Sequence[a] and Sequence[b], which are respectively determined by the initial strategies $x_1(0)$ and $x_2(0)$, converge to limiting strategies. However, under several conditions, the two sequences do not converge to the same limiting strategies, and then SMBRD does not converge. On the other hand, because an AMBRD corresponds to exactly one of the two sequences, the convergence of AMBRD is guaranteed. So for two-player COGs we obtain the following theorem regarding the convergence property of best-response dynamics.
\begin{thm}[Convergence of two-player dynamics]
\label{thm:convergence of SMBRD}
For a two-player COG, SMBRD is guaranteed to converge to an NE starting from an arbitrary $\mathbf{x}(0)$ if and only if the COG has only one NE; in contrast, AMBRD always converges to an NE regardless of the number of NEs in the COG.
\end{thm}

The fact that AMBRD always converges for any initial joint strategies is desirable from an engineering perspective. It guarantees that a distributed network protocol designed based on AMBRD does terminate after a sufficient number of updates, avoiding the so-called ``ping-pong effect''.

\subsection{Convergence Analysis for $K$-player COGs with Linear Best-responses}
\label{subsect:k-player-converge}

The analysis of the convergence property, when it comes to more general case in which a COG consists of $K > 2$ players, becomes difficult due to the lack of convenient properties such as potential games\footnote{The COG is not an exact potential game except for very special choices of network parameters, since in general we have $\frac{\partial^2 u_i}{\partial x_i \partial x_j}\neq \frac{\partial^2 u_j}{\partial x_i \partial x_j}$, $i\neq j$ \cite{mackenzie2006game}, and we were unsuccessful in verifying whether it is an ordinal potential game.} or supermodular games.\footnote{Only for two-player COGs it is possible to convert the COGs into supermodular games \cite{mochaourab2009resource}, and in general COGs are not supermodular since $\frac{\partial^2 u_i}{\partial x_i \partial x_j}>0$ does not always hold for all $x_i, x_j$.} In the following, we analyze the convergence property of $K$-player COGs which admit linear best-response functions.

The best-response function (\ref{equ:Kuser best response function}) implies that player $k$ will allocate all its power to its private band when the aggregated interference from other players in the shared band exceeds a threshold. When the interference is not too strong, the best-response functions become linear without saturation. This occurs if
\begin{equation}
\label{eqn:user-k-linear}
\frac{\beta_k}{\alpha+\beta_k}\left[1+\frac{\sum_{j\in \mathcal{K}\backslash \{k\}}P_{j}c_{j,k}(1-x_j)}{P_k c_{k,k}}\right]<1
\end{equation}
holds for each player $k \in \mathcal{K}$. To ensure that the best-response functions are linear given any initial strategies throughout the execution of best-response dynamics, we need (\ref{eqn:user-k-linear}) to hold for any $x_k \in [0, 1]$, $k \in \mathcal{K}$. This consideration hence leads to the conditions
\begin{equation}
\label{equ:weak interference}
\sum_{j\in \mathcal{K}\backslash \{k\}}P_{j}c_{j,k}<\frac{\alpha}{\beta_k}P_k c_{k,k}, \; \forall k\in \mathcal{K}.
\end{equation}

We first establish that under conditions (\ref{equ:weak interference}) a COG has a unique NE.
\begin{thm}
\label{thm:NE unique}
For a $K$-player COG satisfying (\ref{equ:weak interference}), it has a unique NE.
\end{thm}
\begin{IEEEproof}
Under the assumption of (\ref{equ:weak interference}), all the best-response functions are linear, and by rearranging terms we see that the NEs of the COG should satisfy
\begin{equation}
\label{equ:kuser NE}
\mathbf{x^{*}}=A\mathbf{x^{*}}+\mathbf{b},
\end{equation}
where $A$ is a $K \times K$ matrix in which $A(k,j)=-\frac{\beta_k}{\alpha+\beta_k}\frac{P_j c_{j,k}}{P_k c_{k,k}}$ for $j \neq k$ and $A(k,k)=0$, and $\mathbf{b}(k)=\frac{\beta_k}{\alpha+\beta_k} \frac{\sum_{j\in \mathcal{K}}P_j c_{j,k}}{P_k c_{k,k}}$, for $k = 1, \ldots, K$. From (\ref{equ:weak interference}), we have
\begin{eqnarray}
\label{eqn:row-sum-bound}
\sum_{j \neq k} |A(k, j)| &=& \frac{\beta_k}{\alpha + \beta_k}\frac{1}{P_k c_{k, k}} \sum_{j \neq k} P_j c_{j, k}\nonumber\\
&<& \frac{\beta_k}{\alpha + \beta_k}\frac{1}{P_k c_{k, k}} \frac{\alpha}{\beta_k}P_k c_{k, k}\nonumber\\
&=& \frac{\alpha}{\alpha + \beta_k} \leq 1,
\end{eqnarray}
for all $k = 1, \ldots, K$. Hence from Gershgorin's circle theorem (see, e.g., \cite{golub96:book}), the bound (\ref{eqn:row-sum-bound}) implies that the maximum eigenvalue of matrix $A$ satisfies $|\lambda_{\mathrm{max}}| < 1$. Hence the matrix $I-A$ is nonsingular, so that the solution of $\mathbf{x}^{*}=A\mathbf{x}^{*}+\mathbf{b}$ is unique, given by $\mathbf{x}^{*}=(I-A)^{-1}\mathbf{b}$ with $\mathbf{x}^{*}\in (0,1)^K$ due to Theorem \ref{thm:NE existence}.
\end{IEEEproof}

The convergence of SMBRD is a direct convergence of Theorem \ref{thm:NE unique}.
\begin{thm}
\label{thm:convergence of kuser SMBRD}
For a $K$-player COG satisfying (\ref{equ:weak interference}), SMBRD is guaranteed to converge to the unique NE.
\end{thm}
\begin{IEEEproof}
The updating process of SMBRD can be written as the following iteration,
\begin{equation}
\label{equ:kuser SM updates}
\mathbf{x}(t+1)=A\mathbf{x}(t)+\mathbf{b}.
\end{equation}
From the proof of Theorem \ref{thm:NE unique}, we see that the iteration (\ref{equ:kuser SM updates}) is globally asymptotically stable since all eigenvalues of matrix $A$ satisfy $|\lambda_k|<1$. So we conclude that under (\ref{equ:weak interference}), SMBRD is guaranteed to converge to the unique NE given by $\mathbf{x}^{*}=(I-A)^{-1}\mathbf{b}$.
\end{IEEEproof}

Establishing the convergence of AMBRD is similar while a little more involved, as provided by the proof of the following theorem.
\begin{thm}
\label{thm:convergence of kuser AMBRD}
For a $K$-player COG satisfying (\ref{equ:weak interference}), AMBRD is guaranteed to converge to the unique NE.
\end{thm}
\begin{IEEEproof}
Under the assumption of (\ref{equ:weak interference}), when player $k$ updates its strategy, the updating process of AMBRD can be written as
\begin{equation}
\label{equ:kuser AM updates}
\mathbf{x}(t+1)=A_k \mathbf{x}(t)+\mathbf{b}_k,
\end{equation}
where $A_k$ is a $K \times K$ unit diagonal matrix, except that its $k$-th row is replaced by the $k$-th row of the matrix $A$. The elements of the length-$K$ vector $\mathbf{b}_k$ are all zero except that its $k$-th element is the $k$-th element of $\mathbf{b}$. So if we view a full updating cycle from player $1$ to player $K$ as a whole, the updating iteration is like
\begin{eqnarray}
\label{eqn:AMBRD-iteration}
\mathbf{x}((i+1)K) = \prod_{k = K}^1 A_k \mathbf{x}(iK) + \tilde{\mathbf{b}}, \quad i = 0, 1, \ldots,
\end{eqnarray}
where
\begin{eqnarray}
\tilde{\mathbf{b}} = \prod_{k = K}^2 A_k \mathbf{b}_1 + \prod_{k = K}^3 A_k \mathbf{b}_2 + \ldots + A_K \mathbf{b}_{K - 1} + \mathbf{b}_K.
\end{eqnarray}
So in order to prove that the AMBRD converges, it suffices to establish that all the eigenvalues of $\prod_{k = K}^1 A_k$ satisfy $|\lambda_k| < 1$. For this, we again utilize Gershgorin's circle theorem, showing that the row norm of $\prod_{k = K}^1 A_k$ is smaller than one.

Denote the $k$-th row elements of $A_k$ by $[a_{k, 1}, a_{k, 2}, \ldots, a_{k, K}]$, in which $a_{k, k} = 0$ and $a_{k, j} = A(k, j) = -\frac{\beta_k}{\alpha + \beta_k} \frac{P_j c_{j, k}}{P_k c_{k, k}}$ for $j \neq k$. Let us trace the calculation of $\prod_{k = K}^1 A_k$ to check that all of its absolute row sums are smaller than one. For this, we show by induction the following claim: each of the first $l$ rows of $A^{(l)} = \prod_{k = l}^1 A_k$ has its absolute sum smaller than one, for $l = 1, \ldots, K$.

For $l = 1$, the claim apparently holds. For $l = 2$, $A^{(2)} = A_2 A_1$. Its first row remains $[a_{1,1}, a_{1, 2}, \ldots, a_{1, K}]$, whose absolute row sum is smaller than one, by the assumption of (\ref{equ:weak interference}). Its second row is
\begin{eqnarray*}
\left[a_{2, 1} a_{1, 1}, \; (a_{2, 1} a_{1, 2} + a_{2, 2}), \; (a_{2, 1} a_{1, 3} + a_{2, 3}), \; \ldots, (a_{2, 1} a_{1, K} + a_{2, K})\right],
\end{eqnarray*}
whose absolute row sum is
\begin{eqnarray*}
|a_{2, 1} a_{1, 1}| + \sum_{j = 2}^K |a_{2, 1} a_{1, j} + a_{2, j}| &\leq& |a_{2, 1}| \sum_{j = 1}^K |a_{1, j}| + \sum_{j = 2}^K |a_{2, j}|\\
&<& |a_{2, 1}| \cdot 1 + \sum_{j = 2}^K |a_{2, j}| = \sum_{j = 1}^K |a_{2, j}| < 1.
\end{eqnarray*}
So the claim holds for $l = 2$. Now assume that the claim holds up till $l$, and examine $A^{(l + 1)}$. For notational simplicity we denote the $i$-th row elements of $A^{(l)}$ by $[a^{(l)}_{i, 1}, a^{(l)}_{i, 2}, \ldots, a^{(l)}_{i, K}]$. Since $A^{(l+1)} = A_{l+1} A^{(l)}$, we see that its first $l$ rows remain those of $A^{(l)}$, thus satisfying their absolute row sums smaller than one by assumption. For its $(l+1)$-th row, the first $l$ elements are
\begin{eqnarray*}
a^{(l + 1)}_{l+1, j} = a_{l + 1, 1} a^{(l)}_{1, j} + a_{l + 1, 2} a^{(l)}_{2, j} + \ldots + a_{l + 1, l} a^{(l)}_{l, j}, \quad j = 1, \ldots, l,
\end{eqnarray*}
and the last $K - l$ elements are
\begin{eqnarray*}
a^{(l + 1)}_{l+1, j} = a_{l + 1, 1} a^{(l)}_{1, j} + a_{l + 1, 2} a^{(l)}_{2, j} + \ldots + a_{l + 1, l} a^{(l)}_{l, j} + a_{l+1, j}, \quad j = l + 1, \ldots, K.
\end{eqnarray*}
So we have
\begin{eqnarray*}
\sum_{j = 1}^K \left|a^{(l+1)}_{l+1, j}\right| &\leq& \sum_{i = 1}^l |a_{l+1, i}| \sum_{j = 1}^K \left|a^{(l)}_{i, j}\right| + \sum_{j = l+1}^K |a_{l+1, j}|\\
&<& \sum_{i = 1}^l |a_{l+1, i}| + \sum_{j = l+1}^K |a_{l+1, j}| < 1,
\end{eqnarray*}
where the first inequality is from $|x + y| \leq |x| + |y|$, the second inequality is from the assumption of induction for $A^{(l)}$, and the third inequality is from the assumption for $A_{l + 1}$. As we let $l$ increase from $1$ to $K - 1$, we establish the claim that each of the rows of $A^{(K)} = \prod_{k = K}^1 A_k$ has its absolute sum smaller than one, and thus Gershgorin's circle theorem guarantees that all the eigenvalues of $\prod_{k = K}^1 A_k$ satisfy $|\lambda_k| < 1$.

Consequently, the iteration (\ref{eqn:AMBRD-iteration}) has a unique fixed point as
\begin{eqnarray}
\label{eqn:fixed-point-AMBRD}
\mathbf{x}^\ast = \left(I - \prod_{k = K}^1 A_k\right)^{-1} \tilde{\mathbf{b}},
\end{eqnarray}
which, from the nature of AMBRD, is a NE of the underlying COG. On the other hand, Theorem \ref{thm:NE unique} indicates that the NE is unique under (\ref{equ:weak interference}). Therefore we see that the fixed point in (\ref{eqn:fixed-point-AMBRD}) has to coincide with the unique NE of the COG given in Theorem \ref{thm:NE unique}, i.e., $\mathbf{x}^{*}=(I-A)^{-1}\mathbf{b}$.
\end{IEEEproof}

\section{\textsc{Numerical Results}}
\label{sect:simulation}

In this section, we perform some numerical simulations to illustrate the analysis in Section \ref{sect:convergence}. First, We fix the spectrum allocation as $\alpha=0.5$, $\beta_k={0.5}/{K}$, $k = 1, \ldots, K$. The average power budgets for all the WSPs are identical as $P_k=1$, and the power spectral density of white Gaussian noise is $n_0=10^{-2}$. We use Monte Carlo simulation, in which the initial joint strategies are uniformly distributed in the entire product strategy space, to verify the relationship between the network parameters and the best-response dynamics' convergence property. In simulation, we view an iteration as converged when either the condition
\begin{equation}
\max_{k\in \mathcal{K}} \{|x_k(t+1)-x_k(t)|\}\leq \epsilon = 10^{-2}
\end{equation}
is met for SMBRD, or the condition
\begin{equation}
\max_{k\in \mathcal{K}} \{|x_k(t+j)-x_k(t+j-K)|\}\leq \epsilon = 10^{-2},\;\forall j  = 1,\cdots, K
\end{equation}
is met for AMBRD. The maximum number of updates in an iteration is set as $100$.

\begin{figure}[!ht]
\centering
\includegraphics[width=3.3in,
height=3in]{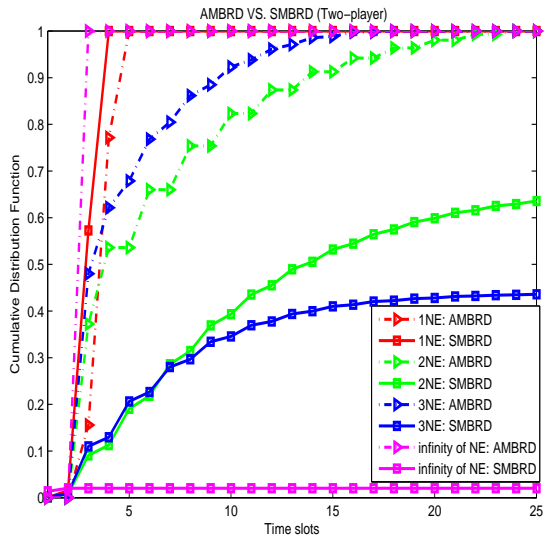}
\caption{Convergence property of two-player COGs}
\label{2user}
\end{figure}

When a COG has only two players, the convergence property of SMBRD and AMBRD by simulations is depicted in Figure \ref{2user}. We consider four sets of network parameters: $({c_{1,2}}/{c_{2,2}},{c_{2,1}}/{c_{1,1}})=(0.4,0.6),(3,4),(3.5,4),(3,3)$, which correspond to four different cases: a COG having one NE, two NEs, three NEs and an infinite number of NEs. We use empirical cumulative distribution functions to characterize each best-response dynamic's convergence property. From Figure \ref{2user}, we verify that AMBRD always converges to an NE for all cases, and that SMBRD only converges when a COG has a unique NE. Nevertheless, when a COG has a unique NE, SMBRD converges more quickly than AMBRD.

\begin{figure}[!ht]
\centering
\includegraphics[width=3.3in,
height=3in]{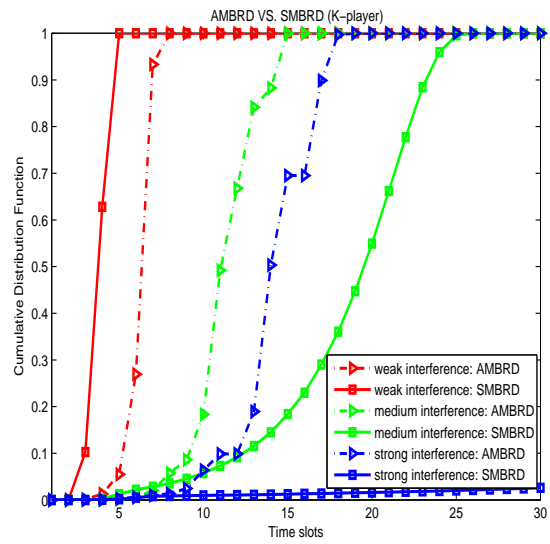}
\caption{Convergence property of four-player COGs}
\label{4user}
\end{figure}

When a COG has four players, we perform a corresponding simulation, with the convergence property depicted in Figure \ref{4user}. We consider three different interference matrices:
\begin{eqnarray*}
\mathbf{C}_1 = \left[
\begin{array}{cccc}
1 & 0.2 & 0.1 & 0.4 \\
0.4 & 1 & 0.5 & 0.3 \\
0.3 & 0.4 & 1 & 0.6 \\
0.4 & 0.2 & 0.5 & 1 \\
\end{array}
\right], \quad
\mathbf{C}_2 = \left[
\begin{array}{cccc}
1 & 0.6 & 1.4 & 1.6 \\
1.4 & 1 & 0.9 & 1.4 \\
2.3 & 1.4 & 1 & 2.0 \\
0.9 & 0.7 & 1.4 & 1 \\
\end{array}
\right], \quad
\mathbf{C}_3 = \left[
\begin{array}{cccc}
1 & 1.4 & 2.0 & 0.9 \\
0.4 & 1 & 1.6 & 2.1 \\
1.4 & 2.2 & 1 & 0.9 \\
1.2 & 2.1 & 3.0 & 1 \\
\end{array}
\right].
\end{eqnarray*}
They respectively correspond to weak interference (conditions (\ref{equ:weak interference}) satisfied for all players), medium interference (conditions (\ref{equ:weak interference}) satisfied for all but one players), and strong interference (conditions (\ref{equ:weak interference}) unsatisfied for all players). From the simulation results, we verify the validity of the analysis in Section \ref{subsect:k-player-converge} for the case of weak interference. Furthermore, we observe that SMBRD and AMBRD may also converge to an NE even when the best-response functions are nonlinear with saturation, although our analysis in Section \ref{sect:convergence} is not able to ensure so. When the interference is weak, SMBRD converges more quickly than AMBRD; whereas as the interference becomes strong, AMBRD converges more quickly than SMBRD.

\section{Conclusion}
\label{sect:conclusion}

In this paper, motivated by the emerging use case of capacity offload, we considered the interference management problem in which different WSPs allocate their transmission power resources between their own private bands and a shared band which is simultaneously available to all of the WSPs. Taking into account the non-cooperate relationship among the WSPs, we formulated the problem into a non-cooperative game and analyzed its properties. We further proposed two distributed learning dynamics for each WSP to individually learn from its local measurement to reach an NE, and analyzed the convergence properties of the dynamics. A number of topics may be explored for future research, including establishing the convergence properties for general $K$-user COGs without linear best-responses, cooperative game-theoretic formulations, and design of effective mechanisms for improved overall utilities for WSPs and even spectrum allocators.

\balance
\bibliographystyle{IEEEtran}
\bibliography{reference}

\begin{thebibliography}{10}
\providecommand{\url}[1]{#1}
\csname url@samestyle\endcsname
\providecommand{\newblock}{\relax}
\providecommand{\bibinfo}[2]{#2}
\providecommand{\BIBentrySTDinterwordspacing}{\spaceskip=0pt\relax}
\providecommand{\BIBentryALTinterwordstretchfactor}{4}
\providecommand{\BIBentryALTinterwordspacing}{\spaceskip=\fontdimen2\font plus
\BIBentryALTinterwordstretchfactor\fontdimen3\font minus
  \fontdimen4\font\relax}
\providecommand{\BIBforeignlanguage}[2]{{%
\expandafter\ifx\csname l@#1\endcsname\relax
\typeout{** WARNING: IEEEtran.bst: No hyphenation pattern has been}%
\typeout{** loaded for the language `#1'. Using the pattern for}%
\typeout{** the default language instead.}%
\else
\language=\csname l@#1\endcsname
\fi
#2}}
\providecommand{\BIBdecl}{\relax}
\BIBdecl

\bibitem{meshkati2009mobility}
F.~Meshkati, Y.~Jiang, L.~Grokop, S.~Nagaraja, M.~Yavuz, and S.~Nanda,
  ``{Mobility and capacity offload for 3G UMTS femtocells},'' in \emph{Proc. IEEE Global
  Telecommunications Conference (GLOBECOM)}, Honolulu, HI, USA, pp. 1--7, Nov.-Dec. 2009.

\bibitem{roh2011femtocell}
J.~Roh, Y.~Ji, Y.~G.~Lee, and T.~Ahn, ``Femtocell traffic offload scheme for core
  networks,'' in \emph{Proc. IFIP International Conference on New Technologies, Mobility and Security (NTMS)}, Paris, France, pp. 1--5, Feb. 2011.

\bibitem{fcc2010rules}
Federal Communications Commission, \emph{Unlicensed Operation in the TV Broadcast Bands}, 47 CFR Part 15, Federal Register, vol. 74, no. 30, Feb. 2009.

\bibitem{nguyen2011impact}
T.~Nguyen, H.~Zhou, R.~A.~Berry, M.~L.~Honig, and R.~Vohra, ``The impact of
  additional unlicensed spectrum on wireless services competition,'' in
  \emph{Proc. IEEE Symposium on New Frontiers in Dynamic Spectrum Access Networks (DySPAN)}, Aachen, Germany, pp. 146--155, Apr. 2011.

\bibitem{peha2009sharing}
J.~Peha, ``Sharing spectrum through spectrum policy reform and cognitive
  radio,'' \emph{Proceedings of IEEE}, vol.~97, no.~4, pp. 708--719, 2009.

\bibitem{han11:book}
Z. Han, D. Niyato, W. Saad, T. Basar, and A. Hjorungnes, \emph{Game Theory in Wireless and Communication Networks: Theory, Models and Applications}, Cambridge University Press, 2011.

\bibitem{belmega2009resource}
E.~V.~Belmega, B.~Djeumou, and S.~Lasaulce, ``Resource allocation games in
  interference relay channels,'' in \emph{Proc. International Conference on Game Theory for Networks (GameNets)}, Istanbul, Turkey, pp. 575--584, May 2009.

\bibitem{mochaourab2009resource}
R.~Mochaourab and E.~Jorswieck, ``Resource allocation in protected and shared
  bands: uniqueness and efficiency of Nash equilibria,'' in \emph{Proc.
  ICST/ACM International Workshop on Game Theory in Communication Networks (Gamecomm)}, Pisa, Italy, pp. 1--10, Oct. 2009.

\bibitem{nash1950equilibrium}
J.~Nash, ``Equilibrium points in $n$-person games,'' \emph{Proceedings of
  National Academy of Science}, vol.~36,
  no.~1, pp. 48--49, 1950.

\bibitem{fudenberg1998theory}
D.~Fudenberg and D.~Levine, \emph{{The Theory of Learning in Games}}, MIT Press, 1998.

\bibitem{fudenberg1991game}
D.~Fudenberg and J.~Tirole, \emph{{Game Theory}}, MIT Press, 1991.

\bibitem{rosen1965existence}
J.~Rosen, ``{Existence and uniqueness of equilibrium points for concave
  $n$-person games},'' \emph{Econometrica: Journal of the Econometric Society},
  vol. 33, no. 3, pp. 520--534, Jul. 1965.

\bibitem{mackenzie2006game}
A.~MacKenzie and L.~DaSilva, ``Game theory for wireless engineers,''
  \emph{Synthesis Lectures on Communications}, vol.~1, no.~1, pp. 1--86, 2006.
  
\bibitem{golub96:book}
G.~H. Golub and C.~F. van Loan, \emph{Matrix Computations}, Johns Hopkins University Press, 1996.

\end{thebibliography}

\newpage

\end{document}